\documentclass[acmsmall]{acmart}

\usepackage{times}
\usepackage{soul}
\usepackage{url}
\usepackage[utf8]{inputenc}
\usepackage{caption}
\usepackage{graphicx}
\usepackage{amsmath}
\usepackage{amsthm}
\usepackage{booktabs}
\usepackage{algorithm}
\usepackage[switch]{lineno}
\usepackage{color}
\usepackage{mathtools}
\usepackage{algpseudocodex}
\usepackage{bm}
\usepackage{subcaption}
\usepackage{xspace}
\usepackage{natbib}
\usepackage{array}
\usepackage{enumitem}

\newlength{\tablewidth}
\setcopyright{none}
\renewcommand\footnotetextcopyrightpermission[1]{}

\newcolumntype{C}[1]{>{\centering\let\newline\\\arraybackslash\hspace{0pt}}m{#1}}

\newcommand{\mech}{\textsc{LendRecoup}\xspace}

\theoremstyle{plain} % default
\newtheorem{theorem}{Theorem}
\newtheorem{lemma}{Lemma}
\newtheorem{proposition}{Proposition}
\newtheorem{corollary}{Corollary}
\theoremstyle{definition}
\newtheorem{definition}{Definition}

\newtheorem*{theorem*}{Theorem}
\newtheorem*{lemma*}{Lemma}
\newtheorem*{proposition*}{Proposition}
\newtheorem*{corollary*}{Corollary}

\begin{document}

\title{Credit Fairness: Online Fairness in Shared Resource Pools}

\author{Seyed Majid Zahedi}
\authornote{Both authors contributed equally to this work.}
\email{smzahedi@uwaterloo.ca}
\affiliation{
       \institution{University of Waterloo}
       \city{Waterloo}
       \country{Canada}
}

\author{Rupert Freeman}
\authornotemark[1]
\email{FreemanR@darden.virginia.edu}
\affiliation{
       \institution{University of Virginia}
       \city{Charlottesville}
       \country{USA}
}

\begin{abstract}
We study repeated allocation of shared resources among agents with time-varying demands and capped linear utilities. In this setting, independently maximizing the minimum endowment-normalized utility in each round satisfies sharing incentives (agents weakly prefer participating in the mechanism to not participating), strategyproofness (agents have no incentive to misreport their demands), and Pareto efficiency. However, this max--min mechanism can lead to large disparities in the total resources received by agents, even when they have the same average demand. We introduce \textit{credit fairness}, a property that, together with Pareto efficiency, strengthens sharing incentives by giving agents who lend resources in early rounds priority toward recouping those resources in later rounds. Credit fairness can be achieved in conjunction with either Pareto efficiency or strategyproofness individually, but we show that, under anonymity, it cannot be achieved together with both. We propose a mechanism that is credit fair and Pareto efficient, and evaluate it in a computational resource-sharing setting.
\end{abstract}

\pagestyle{fancy}
\fancyhead{}
\fancyfoot{}

\maketitle
\thispagestyle{empty}

\section{Introduction}\label{sec:introduction}

In this paper, we consider a set of agents who contribute resources to a shared pool of homogeneous resources.
Each agent owns a fixed endowment but faces time-varying demands.
In such settings, pooling resources naturally increases utilization by accommodating agent-level demand spikes~\citep{kelly1998rate}.

Our primary motivation is the sharing of computing resources~\citep{ghodsi2011dominant,stoica2003core}, such as supercomputers for scientific computing, network bandwidth, and datacenters for internet services.
More broadly, our model accommodates a wide range of applications, including shared energy systems (e.g., microgrids), transportation infrastructure (e.g., airport runways), research infrastructure (e.g., telescopes), and organizational resources shared across departments or units within a company.

In all of these settings, several requirements naturally arise as fundamental to system stability and success.
\emph{Pareto efficiency (PE)} requires that it be impossible to make some agent better off without making another agent worse off, and is essential for any system whose central goal is maximizing resource utilization.
\emph{Sharing incentives (SI)} requires that agents prefer participating in the shared system over relying solely on their own resources.
SI serves as both a fairness property and a stability guarantee: without SI, systems risk collapse as agents opt out.
Finally, \emph{strategyproofness (SP)} requires that agents maximize their utility by truthfully reporting their demands to the system.
Because system performance can only be optimized with respect to reported demands, any discrepancy between reports and true demands can jeopardize efficiency.

While SI is a compelling property,~\citet{vuppalapati2023karma} point out that system fairness may fail even for mechanisms that satisfy SI\@.
Focusing on resource utilization, they observe that the classic \emph{static max-min fairness (SMMF)} mechanism---which lexicographically maximizes the minimum endowment-normalized utility in every round---can lead to large disparities in agent utilities, despite satisfying all aforementioned properties.
Intuitively, this occurs because SMMF is memoryless and therefore does not prioritize agents who contributed resources to others in earlier rounds.
As a result, agents may lend resources to the system without ever being repaid.
To address this issue,~\citet{vuppalapati2023karma} propose \emph{Karma}, which guarantees agents a fraction $\alpha$ of their endowments in each round and allocates the remaining resources by prioritizing agents with low cumulative allocations.

Our goal in this work is to formally capture, via an axiomatic property, the sense in which SMMF is unfair.
To this end, we introduce \emph{credit fairness}.
Intuitively, credit fairness posits the existence of an accounting framework (a ``credit system'') that tracks lending and borrowing and ensures that agents who are net donors are not deprioritized in favor of agents who are net borrowers.
Moreover, when combined with PE, credit fairness constitutes a strengthening of SI: it preserves the stability interpretation of SI while enhancing it from a fairness perspective.
This strengthening comes at a cost: under anonymity, credit fairness is incompatible with simultaneously satisfying PE and SP.

To our knowledge, all previously proposed Pareto-efficient mechanisms in this setting violate credit fairness, including Karma.
We design a simple mechanism, \mech, that is both credit fair and PE\@.
Additionally, we show that \mech is \emph{online strategyproof (OSP)}~\citep{aleksandrov2019strategy}, 
which assumes that agents do not account for future rounds when strategizing.
While this is a relatively weak guarantee, it is, in some sense, the strongest one can hope for, as our impossibility result rules out full SP\@.
Table~\ref{tab:mechanism-properties} summarizes the properties satisfied by existing mechanisms.%
\footnote{Note that DMMF corresponds to Karma with parameter $\alpha = 0$.
For Karma with $0 < \alpha < 1$, we report ``Yes'' when the property holds for all values of $\alpha$ and ``No'' when there exists some $\alpha$ for which the property is violated. Since no formal theoretical analysis exists for Karma with $\alpha > 0$, we report only properties that are straightforward to verify without proof; whether Karma satisfies OSP remains an open question.}

Finally, using data from a real-world environment, we show that \mech{}'s performance is comparable to or better than that of other mechanisms across a range of fairness and efficiency metrics.
Combined with the formal guarantee of credit fairness, these results suggest that \mech is a compelling candidate for practical deployment.

\begin{table}[t!]
	\centering
	\footnotesize
	\begin{tabular}{lccccc}
		\toprule
		\textbf{Mechanism} & \textbf{PE} & \textbf{SP} & \textbf{OSP} & \textbf{SI} & \textbf{Credit Fair} \\
		\midrule
		SMMF  & Yes & Yes & Yes& Yes & No \\
		DMMF  & Yes & No  & Yes & No  & No \\
		Karma ($\alpha \in (0,1))$ & Yes & No & ? &No  & No \\
		Karma ($\alpha=1$) & Yes & No  & ? & Yes & No \\
		$\mech$ & Yes & No  & Yes & Yes & Yes \\
		\bottomrule
	\end{tabular}
	\caption{\textbf{Summary of properties} satisfied by static max-min fairness, dynamic max-min fairness, Karma, and \mech.}
	\label{tab:mechanism-properties}
\end{table}

\section{Related Work}
\label{sec:related-works}

In recent years, there has been renewed interest in repeated allocation mechanisms in dynamic settings where agents have time-varying demands~\citep{freeman2018dynamic,hossain2019sharing,li2021egalitarian,vardi2022dynamic,vuppalapati2023karma,fikioris2024incentives,fikioris2025beyond,cookson2025temporal}.
The work most closely related to ours is that of \citet{freeman2018dynamic}, who study the online allocation of multiple resource units from a shared pool among agents who both contribute to and draw from the pool over multiple rounds.
Agents are assumed to have piecewise-linear utilities that are high up to a demand and low beyond it.
They consider two settings: (I) zero utility beyond demands and (II) strictly positive utility beyond demands.
For (I), they analyze two max--min fair mechanisms: dynamic max--min fair (DMMF),
which lexicographically maximizes the minimum endowment-normalized cumulative utility up to the current round, and static max--min fair (SMMF), which lexicographically maximizes the minimum endowment-normalized utility in the current round.
They show that SMMF satisfies SP, SI, and PE, while DMMF violates SP and SI\@.
For (II), they show that no online mechanism can satisfy PE along with either SI or SP\@.

Another closely related work is that of \citet{vuppalapati2023karma}, who study setting (I) of~\citet{freeman2018dynamic} and propose \emph{Karma}, a mechanism parameterized by $0 \le \alpha \le 1$, which specifies the fraction of each agent's endowment that is guaranteed when demanded.
The remaining resources, together with donations from agents whose demand falls below their guaranteed share, are allocated to agents with excess demand by prioritizing those with the smallest cumulative allocations.
Agents gain credits when others borrow their donated resources and lose credits when borrowing from the pool.
They derive theoretical guarantees only for $\alpha=0$, in which case \emph{Karma} reduces to DMMF, violating SP and SI\@.

Repeated allocation has also been studied in settings where indivisible items arrive over time and must be allocated immediately without revocation.
\citet{zeng2020fairness} and~\citet{benade2024fair} study the trade-off between fairness and efficiency in this setting, while~\citet{aleksandrov2019monotone} examine online fair division under monotonicity constraints. 
\citet{aleksandrov2019strategy} show that, for agents with additive utilities, SP, envy-freeness, and PE cannot be simultaneously achieved against worst-case adversaries.
\citet{benade2018make} provide a deterministic algorithm with vanishing envy over time.
\citet{benade2022dynamic} study a partial information setting, where ordinal but not cardinal information is elicited and values are drawn from i.i.d.\ distributions.
Finally,~\citet{igarashi2024repeated} study the repeated allocation of indivisible goods and chores, providing guarantees on both per-round and cumulative fairness and efficiency.
\section{Preliminaries}\label{sec:preliminaries}
For $k \in \mathbb{N}$, we denote $[k] \coloneqq \{ 1, \ldots, k \}$. Consider a system with $n$ agents and discrete rounds $t=1,2,\ldots$.
(All definitions apply to any finite prefix of rounds.)
At each round, each agent $i$ contributes a fixed \emph{endowment} of $e_i > 0$ units, and the total endowment is $E=\sum_i e_i$. We denote by $E_{-i} = \sum_{j \neq i} e_j$ the total number of units not including $i$'s endowment.
We assume a dynamic setting in which agents have time-varying demands.
At round $t$, agent $i$ has a true demand of $d_{i,t} \ge 0$ units and reports a demand of $d'_{i,t} \ge 0$.
Let $\bm{d}_{i,1:t}=(d_{i,t'})_{1 \le t'  \le t}$ and $\bm{d}'_{i,1:t}=(d'_{i,t'})_{1 \le t'  \le t}$ denote the vectors of agent $i$'s actual and reported demands up to round $t$, respectively.
Further, let $\bm{d}'_{-i, 1:t}$ denote the reported demands of all agents other than $i$ up to round $t$.
At every round $t$, an \emph{online allocation mechanism} $M$ (henceforth simply \emph{mechanism}) assigns each agent an allocation $a^M_{i,t}(\bm{d}'_{1,1:t},\ldots,\bm{d}'_{n,1:t}) = a^M_{i,t}(\bm{d}'_{i,1:t},\bm{d}'_{-i,1:t}) \in \mathbb{R}_{\ge 0}$ with $\sum_{i} a^M_{i,t}(\bm{d}'_{i,1:t},\bm{d}'_{-i,1:t}) \le E$.
We will typically write simply $a_{i,t}$ when the exact mechanism and the demands are clear from context.
Let $\bm{a}^M_{i,1:t}(\bm{d}'_{i,1:t},\bm{d}'_{-i,1:t}) $ (often simply $\bm{a}_{i,1:t}$) denote the vector of agent $i$'s allocations up to round $t$.
Agents derive one unit of utility for every unit of resource they receive up to their demand and do not value additional units; that is, the utility of agent $i$ with demand $d_{i,t}$ at round $t$ is $u_{i,t}(a_{i,t}, d_{i,t}) = \min (d_{i,t}, a_{i,t})$.
We additionally assume that utilities are additive over rounds, so that agent $i$'s total utility over the first $t$ rounds for an allocation $\bm{a}_{i,1:t}$ is $U_{i,t}(\bm{a}_{i,1:t}, \bm{d}_{i,1:t}) = \sum_{t'=1}^t u_{i,t'}(a_{i,t'}, d_{i,t'})$.
For notational simplicity, we write $u_{i,t}(a_{i,t})$ or $U_{i,t}(\bm{a}_{i,1:t})$ when the dependence on $d_{i,t}$ or $\bm{d}_{i,1:t}$ is clear.
It will often be convenient to reason about an agent's utility with respect to their reported, rather than true, demand; for this purpose, we use the shorthand $u'_{i,t}(a_{i,t}) = u_{i,t}(a_{i,t}, d'_{i,t})$ and $U'_{i,t}(\bm{a}_{i,1:t})=U_{i,t}(\bm{a}_{i,1:t}, \bm{d}'_{i,1:t})$.
Although this quantity does not represent an agent's actual utility, it appears often because our definition of credit fairness will disregard incentive considerations and implicitly assume that reported demands equal true demands, as is standard for fairness properties.

Following prior work~\citep{freeman2018dynamic}, we allow the allocations $a_{i,t}$ to be real-valued even though resource units and demands are discrete. We interpret real-valued allocations as being implemented via randomization where agent $i$ is allocated $a_{i,t}$ units in expectation.

\paragraph{Properties of Mechanisms.}
In addition to the new property that we will define later, we focus on three fundamental desiderata: Pareto efficiency (PE), sharing incentives (SI), and strategyproofness (SP).
PE requires that if any agent receives fewer units than their demand, then all units must be allocated to agents who derive positive utility from them, i.e., the sum of agents' utilities equals $E$.
 
 \begin{definition}
 	A mechanism $M$ is \emph{Pareto efficient} if, for every round $t$ and every report profile $(\bm{d}'_{1,1:t},\ldots,\bm{d}'_{n,1:t})$, the following holds: whenever there exists an agent $i$ such that $a^M_{i,t}(\bm{d}'_{i,1:t},\bm{d}'_{-i,1:t}) < d'_{i,t}$, then it must be the case that:
 	\[
 	\sum_{j} \min\!\left(a^M_{j,t}(\bm{d}'_{j,1:t},\bm{d}'_{-j,1:t}), \; d'_{j,t} \right) = E.
 	\]
 \end{definition}
 
 SI requires that agents receive at least as much utility from (truthfully) participating in the system as they would have received by not participating.
 
 \begin{definition}
 	A mechanism $M$ satisfies \emph{sharing incentives} if, for every round $t$, every agent $i$, all true demands $\bm{d}_{i,1:t}$, and any report profile $\bm{d}'_{-i,1:t}$ of the other agents, it holds that:
 	\[
 	U_{i,t}\!\left(\bm{a}^M_{i,1:t}(\bm{d}_{i,1:t}, \bm{d}'_{-i,1:t})\right) \;\ge\; U_{i,t}(\bm{e}_i),
 	\]
 	where $\bm{e}_i = (e_i, \ldots, e_i)$ denotes the allocation in which agent $i$ receives their endowment in every round.
 \end{definition}
 
SP requires that agents never benefit from misreporting.

\begin{definition}
	A mechanism $M$ is \emph{strategy-proof} if, for every agent $i$, every round $t$, all true demands $\bm{d}_{i,1:t}$, all report profiles $\bm{d}'_{i,1:t}$ of agent $i$, and any report profile $\bm{d}'_{-i,1:t}$ of the other agents, it holds that:
	\[
	U_{i,t}\!\left(\bm{a}^M_{i,1:t}(\bm{d}_{i,1:t}, \bm{d}'_{-i,1:t})\right) \;\ge\; U_{i,t}\!\left(\bm{a}^M_{i,1:t}(\bm{d}'_{i,1:t}, \bm{d}'_{-i,1:t})\right).
	\]
\end{definition}

Finally, \emph{online strategyproofness (OSP)}~\citep{aleksandrov2019strategy} requires that, fixing reports of all agents in rounds $1, \ldots, t-1$, no agent can increase their utility in round $t$ by misreporting.

\begin{definition}\label{def:osp}
	A mechanism $M$ is \emph{online strategy-proof} if, for every agent $i$, every round $t$, all demands $\bm{d}_{i,1:t}$, all report profiles $\bm{d}'_{i,1:t}$ of agent $i$ such that $\bm{d}_{i,1:t-1}=\bm{d}'_{i,1:t-1}$, and any report profile $\bm{d}'_{-i,1:t}$ of the other agents, it holds that:
	\[
	u_{i,t}\left(a^M_{i,t}(\bm{d}_{i,1:t},\bm{d}'_{-i,1:t})\right) \;\ge\; u_{i,t}\left(a^M_{i,t}(\bm{d}'_{i,1:t},\bm{d}'_{-i,1:t})\right).
	\]
\end{definition}
\section{Credit Fairness}
\label{sec:credit-fairness}

It is known that 
SMMF satisfies PE, SI, and SP~\citep{freeman2018dynamic}.
However, as~\citet{vuppalapati2023karma} note, SMMF can produce highly inequitable allocations in terms of the number of resources received—as large as a factor of $\Omega(n)$—even between users with the same average demand.
We illustrate this shortcoming with the following example, which is heavily inspired by \citeauthor{vuppalapati2023karma}'s proof of the aforementioned result.

\begin{example}
	\label{ex:smmf-violates-cf}
Suppose there are 3 rounds and 3 agents, each with endowment $e_i = 1$, and the following demands:
\[
\begin{array}{c|cccc}
	& \text{$t=1$} & \text{$t=2$} & \text{$t=3$}\\
	\hline
	\text{Agent 1} & 2 & 2 & 2 \\
	\text{Agent 2} & 2 & 2 & 2 \\
	\text{Agent 3} & 0 & 0 & 6. \\
\end{array}
\]
Note that the total demand of all three agents is the same.
Under SMMF, the resulting allocations are as follows:
\[
\begin{array}{c|cccc}
	& \text{$t=1$} & \text{$t=2$} & \text{$t=3$}\\
	\hline
	\text{Agent 1} & 1.5 & 1.5 & 1 \\
	\text{Agent 2} & 1.5 & 1.5 & 1 \\
	\text{Agent 3} & 0 & 0 & 1. \\
\end{array}
\]
Agents 1 and 2 each receive a utility of $4$, while agent 3 receives a utility of $1$.
\end{example}

Agent 3 could argue that, having lent their endowment to other agents in the first two rounds, they should have the right to reclaim those resources in round 3 when they need them.
However, because SMMF is memoryless, it denies this possibility.
In this section, we formally capture the undesirable behavior of SMMF in this example by introducing \emph{credit fairness}, an axiomatic property that reflects the idea that agents should have priority over resources they previously lent.

\subsection{Formal Definition}\label{subsec:formal-definition}

To define credit fairness, we require the notion of a \emph{credit system}. A credit system $(c_{i,t})_{i \in [n], t \in \mathbb{N}}$ assigns a balance of credits to every agent as a function of the report and allocation histories. We denote by
$
c_{i,t}
=
c_{i,t}(\bm{d}'_{1,1:t-1},\ldots,\bm{d}'_{n,1:t-1},
\bm{a}_{1,1:t-1},\ldots,\bm{a}_{n,1:t-1})
\in \mathbb{R}
$
the number of credits that agent $i$ has at the start of round $t$, and require that $c_{i,1}=0$ for every agent $i$. Let
$
\Delta c_{i,t}
\triangleq
c_{i,t+1}(\bm{d}'_{1,1:t},\ldots,\bm{d}'_{n,1:t},
\bm{a}_{1,1:t},\ldots,\bm{a}_{n,1:t})
-
c_{i,t}(\bm{d}'_{1,1:t-1},\ldots,\bm{d}'_{n,1:t-1},
\bm{a}_{1,1:t-1},\ldots,\bm{a}_{n,1:t-1})$.

\begin{definition}
	\label{def:credit-fairness}
	A mechanism $M$ is \emph{credit fair} if there exists a credit system $(c_{i,t})_{i \in [n], t \in \mathbb{N}}$ such that, for every report history and corresponding allocation history generated by $M$, the following properties hold for all agents $i$ and rounds $t$:
	\begin{enumerate}[label=(CF\arabic*), leftmargin=*, labelwidth=3em, labelsep=0.6em, align=left]
		\item $\min(0, e_i - u'_{i,t}(a_{i,t})) \le \Delta c_{i,t} \le \max(0, e_i - u'_{i,t}(a_{i,t}))$;
		\item If $\sum_{j \neq i} u'_{j,t}(a_{j,t}) > E_{-i}$, then $\Delta c_{i,t} \ge \sum_{j \neq i} u'_{j,t}(a_{j,t})- E_{-i}$;
		\item $\sum_{i \in [n]} \Delta c_{i,t} \le 0$;
		\item $a_{i,t} \ge \min(d'_{i,t}, e_i + \min(0, c_{i,t}))$; and
		\item If $a_{j,t} > \max(0, e_{j} + c_{j,t})$ for some agent $j$, then for all agents $i$ it holds that either $a_{i,t} \ge d'_{i,t}$ or $a_{i,t} \ge e_i + c_{i,t}$.
	\end{enumerate}
\end{definition}

Let us unpack this definition. Condition (CF1) states that agents who receive less than $e_i$ utility (informally, agents donating resources to the system) should not lose credits but may gain up to $e_i - u'_{i,t}(a_{i,t})$. Likewise, agents who receive more than $e_i$ utility (informally, borrowers) should not gain credits but may lose up to $u'_{i,t}(a_{i,t}) - e_i$. Agents whose utility equals their endowment neither gain nor lose credits. Although this condition does not require a one-to-one correspondence between credits and resources, it bounds how far credits can diverge from allocations. At a high level, it justifies interpreting positive credits as claims on resources and negative credits as obligations.

Condition (CF2) requires that if an agent $i$ lends resources to others, i.e., the other agents collectively receive utility exceeding the resources they contribute, then agent $i$ must gain at least one unit of credit for each such resource. This captures the idea that agents should be rewarded when they are pivotal to others receiving valued resources.

Condition (CF3) requires that the credit system be deflationary: the total number of credits cannot increase over time. Since credits represent claims on resources, allowing their total to grow would imply that agents could collectively obtain resources without anyone bearing a corresponding cost.

To understand condition (CF4), consider an agent with $d_{i,t} > e_i$. 
If the agent has positive credits, they must receive at least their endowment $e_i$ whenever their demand is at least $e_i$. If the agent has negative credits, they are guaranteed only $e_i + c_{i,t}$. Intuitively, the system may reclaim from the agent up to the number of resources equal to the credit they owe.

Finally, condition (CF5) relies on the notion of a \emph{credit-adjusted endowment}, defined as $e_i + c_{i,t}$, which reflects both the agent's baseline endowment and whether they are a net borrower or lender. The condition requires that if some agent $j$ receives strictly more than their credit-adjusted endowment in a round, then every other agent with sufficient demand must receive at least their own credit-adjusted endowment. Interpreting the credit-adjusted endowment as an agent's ``fair share,'' this ensures that no agent receives more than their fair share while another receives strictly less.

In Appendix~\ref{subsec:necessity}, we show that none of CF1--CF5 is implied by the remaining four conditions.

In summary, credit fairness establishes a memory-based system in which credits act as a ledger of past deviations from an agent's endowment: agents who receive less accumulate limited future claims, while those who receive more incur limited future obligations. Rather than guaranteeing a fixed minimum share, as adopted by mechanisms such as Karma, credit fairness enforces a dynamic notion of equity in which an agent's entitlement adapts to their credit balance. Agents with positive credits are guaranteed at least their endowment, while agents in debt have their guaranteed minimum reduced according to what they owe.

\subsection{Properties}\label{subsec:properties}

We begin with a useful lemma that characterizes credit updates for any credit-fair and Pareto-efficient mechanism in settings where resources are overdemanded.
The proof of this lemma, along with all other omitted proofs in this section, is provided in Appendix~\ref{sec:app-credit-fairness}.

\begin{lemma}\label{lem:pe-forces-delta}
	Fix a round $t$ with $\sum_{j} d'_{j,t} \ge E$. Let $M$ be a mechanism that
	is \emph{credit fair} and \emph{Pareto efficient}. Then, for every agent $i\in[n]$,
	it holds that $\Delta c_{i,t} \;=\; e_i - a_{i,t}$.
\end{lemma}

We now show that, when combined with Pareto efficiency, credit fairness implies sharing incentives. The proof is omitted for space constraints, but the main idea is to show that if an agent receives less than their SI baseline utility $u'_{i,t}(e_i)$ in some round, then that agent must have nonpositive credits, even after the round-$t$ allocations, i.e., $c_{i,t+1} \le 0$. This fact is sufficient to guarantee that the agent's cumulative utility still does not fall below their SI baseline.

\begin{theorem}\label{thm:CF-implies-SI}
	Let $M$ be a credit-fair and Pareto efficient mechanism.
	Then $M$ satisfies sharing incentives.
\end{theorem}

As a corollary of Theorem~\ref{thm:CF-implies-SI}, any Pareto efficient mechanism that violates sharing incentives cannot be credit fair.
Since DMMF is Pareto efficient but violates sharing incentives, it follows that DMMF is not credit fair.

\begin{corollary}
	The dynamic max-min fair mechanism violates credit fairness.
\end{corollary}

Furthermore, we show that SMMF, despite satisfying sharing incentives and Pareto efficiency, violates credit fairness, which further validates the necessity of our definition.

\begin{proposition}
	\label{prop:smm-not-credit-fair}
	The static max-min fair mechanism violates credit fairness.
\end{proposition}

To our knowledge, SMMF is the only known mechanism that satisfies SP, PE, and SI, yet it violates credit fairness. 
This naturally raises the question of whether there exists a mechanism that satisfies the stronger combination of SP, PE, and credit fairness.
Our next result shows that the answer is no, at least under the mild restriction of anonymity. A mechanism is \textit{anonymous} if permuting the agents results in the allocation being permuted in the same way; in other words, the mechanism does not depend on the identities of the agents.
Whether anonymity can be removed from the following theorem is an open question.

\begin{theorem} \label{thm:impossibility}
	No mechanism is anonymous, strategyproof, credit fair, and Pareto efficient.
\end{theorem}

\begin{proof}
	Consider $n=3$ agents, each with endowment of $e_i = 1$.
	We examine the first $5$ rounds with the following demands:
	{\footnotesize
		\[
			 \begin{array}{c|ccccc}
				 & \text{$t=1$} & \text{$t=2$} & \text{$t=3$} & \text{$t=4$}& \text{$t=5$} \\
				 \hline
				 \text{Agent 1} & 1 & 2 & 0 & 0 & 3 \\
				 \text{Agent 2} & 3 & 0 & 1 & 1 & 2 \\
				 \text{Agent 3} & 0 & 2 & 2 & 2 & 0.\\
			 \end{array}
	\]}
	Let $M$ be credit fair, PE, and anonymous (and therefore also satisfies SI). Then $M$ must produce the following allocations:
	{\footnotesize
	\[
		\begin{array}{c|ccccc}
			& \text{$t=1$} & \text{$t=2$} & \text{$t=3$} & \text{$t=4$}& \text{$t=5$} \\
			\hline
			\text{Agent 1} & 1 & 1 & 0 & 0 & 2 \\
			\text{Agent 2} & 2 & 0 & 1 & 1 & 1 \\
			\text{Agent 3} & 0 & 2 & 2 & 2 & 0,\\
		\end{array}
	\]}
	and the following credit balances:
	{\footnotesize
	\[
		\begin{array}{c|ccccc}
			& \text{$t=1$} & \text{$t=2$} & \text{$t=3$} & \text{$t=4$}& \text{$t=5$} \\
			\hline
			\text{Agent 1} &0 & 0 & 0 & 1 & 2 \\
			\text{Agent 2} &0 & -1 & 0 & 0 & 0 \\
			\text{Agent 3} &0 & 1 & 0 & -1 & -2.\\
		\end{array}
	\]}
	Note that, given the demands and allocations, these credit balances are implied by Lemma~\ref{lem:pe-forces-delta}.
	
	The first-round allocations are implied by SI and PE\@.
	In the second round, SI requires that both agents 1 and 3 receive at least one unit, and (CF5) then requires that the additional unit be allocated to agent 3.
	The third- and fourth-round allocations are dictated by PE\@.
	Finally, the fifth-round allocation is implied by (CF5), which does not allow agent 2 to receive more than one unit unless agent 1 receives at least three.
	
	Now suppose agent~1 instead reports a demand of~0 in the first round.
	In this case, the allocations must be:
	{\footnotesize
	\[
		\begin{array}{c|ccccc}
			& \text{$t=1$} & \text{$t=2$} & \text{$t=3$} & \text{$t=4$}& \text{$t=5$} \\
			\hline
			\text{Agent 1}&0 & 1.5 & 0 & 0 & 3 \\
			\text{Agent 2}&3 & 0 & 1 & 1 & 0 \\
			\text{Agent 3}&0 & 1.5 & 2 & 2 & 0,\\
		\end{array}
	\]}
	with the following credits:
	{\footnotesize
	\[
		\begin{array}{c|ccccc}
			& \text{$t=1$} & \text{$t=2$} & \text{$t=3$} & \text{$t=4$}& \text{$t=5$} \\
			\hline
			\text{Agent 1}&0 & 1 & 0.5 & 1.5 & 2.5 \\
			\text{Agent 2}&0 & -2 & -1 & -1 & -1 \\
			\text{Agent 3}&0 & 1 & 0.5 & -0.5 & -1.5 .\\
		\end{array}
	\]}
	
	To verify the allocations, observe that the allocations in the first, third, and fourth rounds are required by PE, the second-round allocation follows from anonymity, and the fifth-round allocation is implied by (CF5).
	As a result of the misreport, agent 1's total utility increases by 0.5, from 4 to 4.5, demonstrating a profitable deviation.
	Therefore, $M$ is not SP.
\end{proof}

While SP, PE, and credit fairness cannot all be satisfied simultaneously by an anonymous mechanism, any two of them can.
SMMF is SP and PE but not credit fair.
The static mechanism that allocates each agent their endowment $a_{i,t}=e_i$
at each round is SP and credit fair (SP is obvious; see Appendix~\ref{sec:app-credit-fairness} for a proof of credit fairness).
In the next section we will introduce a mechanism that is credit fair and PE.
\section{\mech Mechanism}\label{sec:algorithm}
We present a mechanism, \emph{\mech}, that is credit fair and PE\@.
At each round, \mech allocates resources by first satisfying demands as much as possible up to agents' credit-adjusted endowment $\max(0,e_i+c_{i,t})$.
If resources remain after every agent receives $\min (d'_{i,t}, \max(0,e_i+c_{i,t}))$, the surplus is distributed among agents with remaining demand, prioritizing agents with low cumulative allocations relative to their endowments. 
If resources still remain after all agents receive their demands, then remaining resources are allocated to agents with low demand in the current round.
Credits are updated in a one-to-one correspondence with resource transfers: an agent gains (loses) one credit for each unit of resource lent to (borrowed from) the system.

To formally specify \mech, we use the Proportional Sharing With Constraints (\texttt{PSWC}) procedure of~\citet{freeman2018dynamic}.
This procedure takes as input a total resource amount $A$, agent weights $\bm{w}$, and agent-specific lower and upper bounds $\bm{m}$ (for ``minimum'') and $\bm{l}$ (for ``limit'').
It outputs per-agent allocations by finding a scalar $x$ such that each agent $i$ receives $x w_i$ units whenever $m_i \le x w_i \le l_i$; otherwise, agent $i$ receives $m_i$ if $x w_i < m_i$, or $l_i$ if $x w_i > l_i$.
The \texttt{PSWC} procedure can be implemented in $O(n \log n)$ time using the Divvy algorithm~\citep{gulati2012demand}.

Pseudocode for \mech is given in Algorithm~\ref{alg:credit-fair}.
To implement prioritization of agents with low cumulative allocations relative to their endowments, we conceptually ``reallocate'' all resources from previous rounds using the PSWC procedure, with minimum allocations set to the allocations already made. This allows any remaining uncommitted resources at round $t$ to be used to equalize cumulative allocations relative to agents' endowments as much as possible.

\begin{algorithm}[t!]
	\caption{\mech}
	\label{alg:credit-fair}
	\begin{algorithmic}[1]
		\Require Set of agents $[n]$, endowments $\bm{e} = (e_i)_{i}$ (with $E = \sum_{i} e_i$), and online reports $\bm{d}'_t = (d'_{i,t})_{i}$ revealed in each round.
		\Ensure For each round $t$, allocations $\bm{a}_t = (a_{i,t})_{i}$ and credits $\bm{c}_{t+1} = (c_{i,t+1})_{i}$.

		\LComment{Initializing credits to zero}
		\State $\bm{c}_1 \gets \bm{0}$

		\For{$t = 1,2,\ldots$}
			\If{$\sum_{i} d'_{i,t} \le E$} \Comment{No shortage}
				\LComment{Call \texttt{PSWC} with minima at demands}
				\State $\bm{a}_t \gets \texttt{PSWC}\left(A=E, \bm{w}=\bm{e}, \bm{m}=\bm{d}'_t, \bm{l}=\bm{+\infty} \right)$

			\Else \Comment{Shortage}
				\LComment{Cap demands at credit-adjusted endowments}
				\State $\bm{\bar{d}_t} \gets \min (\bm{d}'_t, \max (\bm{0}, \bm{e}+\bm{c}_t))$
				\If{$E \le \sum_{i} \bar{d}_{i,t}$}
					\LComment{Call \texttt{PSWC} up to capped demands}
						\State $\bm{a}_t \gets  \texttt{PSWC}\left(A=E, \bm{w}=\bm{e}, \bm{m}=\bm{0}, \bm{l}=\bm{\bar{d}_t} \right)$
				\Else
					\LComment{Prioritize low endowment-normalized cumulative allocations}
					\State $\bm{m} \gets \sum_{t'=1}^{t-1} \bm{a}_{t'} + \bm{\bar{d}_t}$
					\State $\bm{l} \gets \sum_{t'=1}^{t-1} \bm{a}_{t'} +\bm{d}'_{t}$
					\State $\tilde{\bm{a}}_t \gets \texttt{PSWC}\left(A=tE, \bm{w}=\bm{e}, \bm{m}, \bm{l} \right)$
					\State $\bm{a}_t \gets \tilde{\bm{a}}_t - \sum_{t'=1}^{t-1} \bm{a}_{t'}$
				\EndIf
			\EndIf
			\LComment{Update credits based on lending and borrowing}
			\State $\bm{c}_{t+1} \gets \bm{c}_t + \bm{e} - \bm{a}_t$
		\EndFor
	\end{algorithmic}
\end{algorithm}

\begin{example}
	Consider the instance from Example~\ref{ex:smmf-violates-cf}. On this example, \mech makes the following allocations:
	\[
	\begin{array}{c|cccc}
		& \text{$t=1$} & \text{$t=2$} & \text{$t=3$}\\
		\hline
		\text{Agent 1} & 1.5 & 1.5 & 0 \\
		\text{Agent 2} & 1.5 & 1.5 & 0 \\
		\text{Agent 3} & 0 & 0 & 3, \\
	\end{array}
	\]
	with the following credit balances:
	\[
	\begin{array}{c|cccc}
		& \text{$t=1$} & \text{$t=2$} & \text{$t=3$}\\
		\hline
		\text{Agent 1} & 0 & -0.5 & -1 \\
		\text{Agent 2} & 0 & -0.5 & -1 \\
		\text{Agent 3} & 0 & 1 & 2. \\
	\end{array}
	\]
	In all rounds the algorithm is in the shortage branch (Line 7). The critical round is $t=3$, in which we have credit-adjusted endowments $\bar{d}_{1,3}=\bar{d}_{2,3}=0$ and $\bar{d}_{3,3}=3$, yielding $\sum_i \bar{d}_{i,t} = 3 \ge E=3$. Therefore, \mech runs \texttt{PSWC} with minimum allocations 0 and maximum allocations $\bm{\bar{d}_t}$, resulting in all 3 resources being allocated to agent 3.
\end{example}

\mech is PE, since it never allocates resources that are not valued unless all demands are exhausted.

\begin{proposition}
	\mech is Pareto efficient.
\end{proposition}

We next show that \mech is credit fair.

\begin{theorem}
	\mech satisfies credit fairness.
\end{theorem}

\begin{proof}
	We examine the five conditions individually. 
	
	\medskip
	\noindent
	\textbf{Condition (CF1)} holds in the shortage case by the definition of the credit update and the fact that $a_{i,t} = u'_{i,t}(a_{i,t})$ in this case; hence, $\Delta c_{i,t} = e_i - a_{i,t}$ for all $i$.
	In the no-shortage case, the same argument applies to agents with $a_{i,t} = d'_{i,t}$.
	Otherwise, note that $a_{i,t} > d'_{i,t}$ implies $a_{i,t} \le e_i$, since the \texttt{PSWC} call never reaches $x > 1$ (as $x > 1$ would allocate more than $e_i$ resources to every agent, violating the capacity constraint).
	For such agents, we have $0 \le \Delta c_{i,t} = e_i - a_{i,t} \le e_i - u'_{i,t}(a_{i,t})$.
	
	\medskip
	\noindent\textbf{Condition (CF2).}
	Fix any agent $i$ and round $t$.
	Observe that:
	$\sum_{j \neq i} u'_{j,t}(a_{j,t}) \le \sum_{j \neq i} a_{j,t} = \sum_{j} a_{j,t} - a_{i,t} = E - a_{i,t}$.
	Rearranging and substituting $\Delta c_{i,t} = e_i - a_{i,t}$ gives:
	$\sum_{j \neq i}  u'_{j,t}(a_{j,t}) - E_{-i} \le e_i-a_{i,t} = \Delta c_{i,t}$.

	\medskip
	\noindent\textbf{Condition (CF3)} holds trivially since
	$\sum_{j} \Delta c_{j,t} = \sum_{j} (e_j - a_{j,t}) = E - \sum_j a_{j,t} = 0$.

	\medskip
	\noindent\textbf{Condition (CF4).}
	Fix an agent $i$ and round $t$.
	We consider the following two cases.

	\smallskip
	\noindent\emph{Case 1: $\sum_{i} d'_{i,t} \le E$.}
	In this case, Algorithm~\ref{alg:credit-fair} calls \texttt{PSWC} with minima $\bm m=\bm d'_t$, which implies
	$a_{i,t} \;\ge\; d'_{i,t} \;\ge\; \min\!\bigl(d'_{i,t},\, e_i + \min(0,c_{i,t})\bigr)$,
	as required by (CF4).

	\smallskip
	\noindent\emph{Case 2: $\sum_{i} d'_{i,t} > E$.}
	Define $\kappa_{i,t}\coloneqq \max(0,e_i+c_{i,t})$ so that $\bar{d}_{i,t}\coloneqq \min(d'_{i,t},\kappa_{i,t})$.
	We consider two subcases.

	\smallskip
	\noindent\emph{Case 2a: $E \le \sum_{j}\bar{d}_{j,t}$.}
	The mechanism calls \texttt{PSWC} with weights $\bm w=\bm e$, minima $\bm m=\bm 0$, and upper limits $\bm l=\bm{\bar{d}_t}$.
	Without upper limits, \texttt{PSWC} would return the proportional allocation $a_{i,t}=e_i$ (since $A=E=\sum_{i} e_i$). With the upper limits, \texttt{PSWC} can only reduce the allocation of agents whose proportional share exceeds their cap, and redistribute any leftover to agents whose cap exceeds their proportional share.
	In particular, we have $a_{i,t} \;\ge\; \min(e_i,\bar{d}_{i,t})$ for all agents $i$.
	Now consider $c_{i,t}\ge 0$ and $c_{i,t}<0$ separately.
	If $c_{i,t}\ge 0$, then $\kappa_{i,t}\ge e_i$, hence $\bar{d}_{i,t} = \min(d'_{i,t},\kappa_{i,t}) \ge \min(d'_{i,t}, e_i)$.
	Therefore, we have:
	\begin{equation*}
	a_{i,t} \ge \min(e_i, \bar{d}_{i,t}) \ge \min(e_i,d'_{i,t}) = \min\!\bigl(d'_{i,t},\, e_i+\min(0,c_{i,t})\bigr).
	\end{equation*}
	If $c_{i,t} < 0$ and $e_i + c_{i,t} < 0$, the desired inequality holds trivially since allocations are nonnegative.
	If $c_{i,t} < 0$ and $e_i + c_{i,t} \ge 0$, then $\kappa_{i,t} = e_i + c_{i,t}$, and therefore $\bar{d}_{i,t} = \min(d'_{i,t},e_i + c_{i,t})$.
	This implies $\bar{d}_{i,t} \le e_i$, and hence $\min(e_i, \bar{d}_{i,t}) = \bar{d}_{i,t}$.
	As a result, we have:
	\begin{equation*}
	a_{i,t} \;\ge\; \bar{d}_{i,t} = \min(d'_{i,t},\, e_i + c_{i,t}) = \min\!\bigl(d'_{i,t},\, e_i + \min(0, c_{i,t})\bigr).
	\end{equation*}

	\smallskip
	\noindent\emph{Case 2b: $E > \sum_{j}\bar{d}_{j,t}$.}
	In this case, the mechanism calls \texttt{PSWC} with minima set to $\bm{m} = \sum_{t'=1}^{t-1} \bm{a_{t'}} + \bm{\bar{d}_t}$, which guarantees that the cumulative allocation after round $t$ is at least $\sum_{t'=1}^{t-1}a_{i,t'} + \bar{d}_{i,t}$.
	Thus $a_{i,t} \ge \bar{d}_{i,t} = \min(d'_{i,t},\kappa_{i,t})$.
	If $c_{i,t} \ge 0$, then $\kappa_{i,t} \ge e_i$, which implies $\bar{d}_{i,t} \ge \min(d'_{i,t}, e_i) = \min(d'_{i,t},e_i+\min(0,c_{i,t}))$.
	If $c_{i,t} < 0$ and $e_i+c_{i,t} < 0$, the bound is trivial.
	If $c_{i,t} < 0$ and $e_i+c_{i,t} \ge 0$, then $\kappa_{i,t} = \max(0,e_i+c_{i,t})$, and the same argument as above yields $a_{i,t} \ge \min(d'_{i,t}, e_i + c_{i,t})= \min(d'_{i,t}, e_i + \min (0, c_{i,t}))$.

	\medskip
	\noindent\textbf{Condition (CF5).}
	Fix a round $t$ and suppose there exists an agent $j$ such that:
	$a_{j,t} \;>\; \max(0,e_j+c_{j,t}) \;=\; \kappa_{j,t}$.
	We must show that for every agent $i$, either $a_{i,t} \ge d'_{i,t}$ or $a_{i,t}\ge e_i+c_{i,t}$.

	\smallskip
	\noindent\emph{Case 1: $\sum_i d'_{i,t} \le E$.}
	In this case, the mechanism allocates $a_{i,t} \ge d'_{i,t}$ to all agents,
	so (CF5) holds trivially.

	\smallskip
	\noindent\emph{Case 2: $\sum_i d'_{i,t} > E$ and $E \le \sum_{i} \bar{d}_{i,t}$.}
	Here, \texttt{PSWC} is called with upper limits $\bm{l} = \bm{\bar{d}_t}$.
	Therefore, for every agent $i$ we have $a_{i,t} \le \bar{d}_{i,t} \le \kappa_{i,t}$.
	Therefore, the premise $a_{j,t} > \kappa_{j,t}$ can never hold, and (CF5) holds vacuously.

	\smallskip
	\noindent\emph{Case 3: $\sum_i d'_{i,t} > E$ and $E > \sum_{i}\bar{d}_{i,t}$.}
	In this case, as shown above, the minimum constraints guarantee that every agent $k$ receives at least $\bar{d}_{k,t} = \min(d'_{k,t},\kappa_{k,t})$ units.
	Now fix any agent $i$.
	If $d'_{i,t} \le \kappa_{i,t}$, then $\bar{d}_{i,t}=d'_{i,t}$, and hence $a_{i,t}\ge d'_{i,t}$.
	Otherwise, $d'_{i,t}>\kappa_{i,t}$, in which case $\bar{d}_{i,t}=\kappa_{i,t}$ and
	$a_{i,t}\;\ge\;\bar{d}_{i,t}=\kappa_{i,t}=\max(0,e_i+c_{i,t})\;\ge\;e_i+c_{i,t}$.
	Therefore, for every $i$, either $a_{i,t} \ge d'_{i,t}$ or $a_{i,t}\ge e_i+c_{i,t}$.
\end{proof}

Since \mech is anonymous, Theorem~\ref{thm:impossibility} implies that it cannot be strategyproof.
However, we can show that it satisfies the weaker notion of \emph{online strategyproofness}.

\begin{theorem}\label{thm:mech-osp}
	\mech is online strategy-proof.
\end{theorem}

 The proof of this theorem is provided in Appendix~\ref{sec:app-algorithm}.
\section{Experiments}\label{sec:experiments}

We compare three alternatives
against \mech: (i) static max-min fairness (SMMF), (ii) dynamic max-min fairness (DMMF), and (iii) Karma (with $\alpha = 0.5$). 
For Karma, we choose $\alpha=0.5$ because \citet{vuppalapati2023karma} use that value of $\alpha$ in their own empirical evaluation. Recall that Karma with $\alpha=0$ corresponds to DMMF\@.

\subsection{Workloads}\label{subsec:workloads}
We evaluate mechanisms over 50 independent simulation repetitions.
In each repetition, we simulate a system with 50 agents over 500 rounds.
To generate demand profiles, we use Google cluster traces~\citep{clusterdata:Wilkes2011,clusterdata:Reiss2011}, similar to~\cite{freeman2018dynamic}.
These traces record task-processing events on a 12.5k-machine cluster over roughly one month in May 2011.
Each task submission specifies resource demands requested by the submitting agent, including CPU, memory, and disk space.
Demands are normalized relative to the maximum capacity of each resource across all machines in the traces.
We focus exclusively on CPU requests.

\begin{table*}[t!]
	\centering
	\footnotesize
	\begin{tabular}{C{0.12\tablewidth}C{0.09\tablewidth}C{0.105\tablewidth}C{0.115\tablewidth}C{0.1\tablewidth}C{0.09\tablewidth}C{0.08\tablewidth}C{0.08\tablewidth}}
		\toprule
		\textbf{Mechanism} & \textbf{NW} (std) & \textbf{Min SIx} (std) & \% \textbf{SI Vio.} (std) &
		\textbf{WMM} (std) & \textbf{NMM} (std) & \textbf{WEq} (std) & \textbf{NEq} (std) \\
		\midrule
		\mech &
		1.018 (0.54) & \textbf{1.00} (0.00) & \textbf{0.0} (0) &
		0.26 (0.15) & 0.013 (0.01) & 0.26 (0.15) & \textbf{0.87} (0.09) \\
		DMMF &
		\textbf{1.019} (0.54) & 0.96 (0.04) & 35.8 (12.5) &
		\textbf{0.40} (0.17) & 0.010 (0.01) & \textbf{0.42} (0.19) & 0.84 (0.11) \\
		SMMF &
		1.017 (0.54) & \textbf{1.00} (0.00) & \textbf{0.0} (0) &
		0.08 (0.09) & \textbf{0.018} (0.01) & 0.08 (0.09) & 0.85 (0.09) \\
		Karma &
		\textbf{1.019} (0.54) & 0.96 (0.03) & 36.5 (10.6) &
		0.39 (0.18) & 0.011 (0.01) & 0.41 (0.19) & 0.83 (0.11) \\
		\bottomrule
	\end{tabular}
	\caption{Simulation results for trace-driven demands.}
	\label{tab:all_results_google}
\end{table*}

Following~\cite{freeman2018dynamic}, we divide time into 15-minute rounds and define each agent's demand in a round as the total CPU requested by tasks submitted during that interval.
After processing the traces, we remove agents with constant demands or with average demand below a small threshold.
Each remaining agent is assigned an endowment equal to its average demand across all rounds.
This reflects real-world settings where agents contribute resources roughly proportional to their long-term demand and rely on sharing during demand spikes.

\subsection{Evaluation Metrics}\label{subsec:metrics}
We evaluate mechanisms using agents' cumulative utilities $U_i$.
Let $U_i^{\text{static}}$ denote the utility agent $i$ receives under the \emph{static} allocation, where each agent obtains their endowment in every round.
We consider the following metrics.

\paragraph{Nash Welfare (NW)} We define NW as $\sum_{i} w_i \log(U_i)$, where $w_i = \frac{e_i}{\sum_j e_j}$.
NW balances efficiency and fairness by rewarding high total utility while penalizing inequality.
We report NW normalized by the value achieved under the static allocation.

\paragraph{Sharing Index (SIx)} For agent $i$, the sharing index is $\text{SIx}_i = U_i / U_i^{\text{static}}$.
It measures the individual gain or loss from sharing; values below one indicate harm.
We report the minimum SIx across agents and the percentage of agents with $\text{SIx}_i < 1$.

\paragraph{Weighted Min--Max (WMM)} The WMM ratio is $\frac{\min_i \; U_i / w_i}{\max_i \; U_i / w_i}$.
It measures fairness by comparing the worst- and best-off agents after weight normalization, with larger values indicating greater equity.
This metric can be sensitive to agents with consistently low demands, for whom $U_i/w_i$ will be small under any mechanism.
This motivates the normalized variant below.

\paragraph{Normalized Min--Max (NMM)} The NMM ratio is $\frac{\min_i \; \text{SIx}_i}{\max_i \; \text{SIx}_i}$, capturing inequality in realized performance relative to agents' standalone guarantees.

\paragraph{Weighted Equity (WEq)} We define WEq as $\frac{\min_i \; U_i / w_i}{\text{median}_i \; U_i / w_i}$.
It compares the worst-off agent to a typical agent after weight normalization,%
\footnote{The inverse of this metric appears in~\cite{vuppalapati2023karma}, where it is called \emph{disparity}.}
with larger values indicating greater equity.
WEq is less sensitive to outliers than the min--max ratios above, though still affected by low-demand agents.

\paragraph{Normalized Equity (NEq)} We define NEq as $\frac{\min_i \; \text{SIx}_i}{\text{median}_i \; \text{SIx}_i}$, and is less sensitive than WEq to agents with low demands.

\subsection{Results}

As shown in Table~\ref{tab:all_results_google}, NW is tightly clustered across mechanisms. Since all four mechanisms are Pareto efficient, their utilitarian welfares $\sum_i U_i$ are identical; thus NW reflects only distributional effects in these simulations. Overall, DMMF and Karma achieve the highest values, with \mech extremely close behind, while SMMF trails slightly. The small spread suggests that none of the mechanisms delivers a decisive welfare advantage, and that \mech attains near-frontier performance despite its stronger guarantees.

For SIx, \mech and SMMF stand out by completely avoiding harm from sharing: both report a minimum value of $1.00$ and $0\%$ of agents with $\mathrm{SIx}_i < 1$. In contrast, DMMF and Karma exhibit harm for a substantial minority of agents, with minimum $\mathrm{SIx}_i$ values below one ($0.96$) and roughly a third of agents experiencing losses relative to the constant-allocation baseline ($\approx 36\%$).

These differences matter because $\mathrm{SIx}_i$ directly captures whether sharing can leave participants worse off than not participating. By eliminating observed harm (both in worst case and incidence), \mech offers a stronger practical value proposition than mechanisms achieving similar welfare while shifting costs onto a subset of agents.

For WMM and WEq, DMMF performs best, with Karma nearly tied, while \mech achieves roughly 60\% of DMMF's performance; SMMF trails significantly. In contrast, under the normalized metrics NMM and NEq, Karma and DMMF perform relatively worse: SMMF performs best for NMM followed by \mech, while \mech performs best for NEq. The disparity between the weighted and normalized metrics is notable; understanding the causes of this gap is an interesting direction for future work. Among the mechanisms considered, \mech is the most robust across metrics, again suggesting that it yields a more even distribution of utilities across agents.
\section{Conclusion}\label{sec:conclusion}

We introduce the notion of credit fairness in resource-sharing systems, motivated by the idea that agents who lend resources to the system in early rounds should receive priority later.
Combined with Pareto efficiency, credit fairness strengthens sharing incentives, making it both a fairness and stability notion for shared systems.
Although this strengthening comes at a cost---credit fairness cannot be achieved by an anonymous mechanism simultaneously with both Pareto efficiency and strategyproofness---we introduce \mech, a novel mechanism that is credit fair, Pareto efficient, and online strategyproof.
Experiments on real-world computational resource-sharing traces show that \mech achieves performance comparable to existing state-of-the-art mechanisms.
Together with its formal credit-fairness guarantee, these results position \mech as a compelling candidate for practical deployment.

Our work raises several open questions.
We study a setting with a single resource type, capped linear utility functions, and a fixed set of agents and resource units.
Extending credit fairness to more general settings is an interesting direction for future work.
Even within our setting, the design space of mechanisms and desiderata remains far from fully explored.
For instance, are there mechanisms that achieve stronger theoretical guarantees (e.g., $\epsilon$-Nash SP) and/or improved empirical performance than the four mechanisms studied here?

\bibliographystyle{named}
\bibliography{sections/ref}

\newpage

\appendix
\section{Omitted Material from Section \ref{sec:credit-fairness}}\label{sec:app-credit-fairness}

\subsection{Irredundancy of the Credit-Fairness Conditions}
\label{subsec:necessity}

\begin{proposition}
For each $k\in\{1,\ldots,5\}$, there exists a two-agent demand, allocation, and credit history for which condition (CF$\ell$) holds in every round for every $\ell \neq k$, while condition (CF$k$) fails in at least one round.
Hence, none of CF1--CF5 is pointwise implied by the other four conditions.
\end{proposition}

\begin{proof}
Throughout the proof, there are two agents with endowments $e_1 = e_2 = 1$
and initial credits $(c_{1,1}, c_{2,1}) = (0,0)$.
Thus, $E_{-1}=E_{-2}=1$.

\paragraph{\textbf{CF1 is not implied by CF2--CF5.}}
Consider one round with
\[
	\bm d'_1 = \bm a_1=(1,1), \qquad \text{and} \qquad \Delta\bm c_1=(-1,0).
\]
For each agent, the other agent receives utility exactly $1=E_{-i}$, so the premise of CF2 is false.
CF3 holds because
\[
	\Delta c_{1,1}+\Delta c_{2,1}=-1\le 0.
\]
CF4 holds because each agent receives their full demand, and the premise of CF5 is false because neither agent receives strictly more than their credit-adjusted endowment of $1$.
However, CF1 fails for agent~1.
Since $u'_{1,1}(a_{1,1}) = e_1 = 1$, CF1 requires $\Delta c_{1,1} = 0$, whereas $\Delta c_{1,1} = -1$.

\paragraph{\textbf{CF2 is not implied by CF1, CF3, CF4, and CF5.}}
Consider one round with
\[
	\bm d'_1 = \bm a_1= (0.5,1.5), \qquad \text{and} \qquad \Delta \bm c_1=(0,0).
\]
CF1 holds because the permitted update intervals are $[0,0.5]$ and $[-0.5,0]$,
respectively, and both contain zero.
CF3 holds with equality, and CF4 holds because both agents receive their full demands.

Agent~2 receives strictly more than their credit-adjusted endowment of $1$, but CF5 nevertheless holds because both agents are fully satisfied.
However, CF2 fails for agent~1.
Agent~2 receives utility $1.5>E_{-1}=1$, so CF2 requires $\Delta c_{1,1} \ge 1.5-1 = 0.5$.
Instead, $\Delta c_{1,1}=0$.

\paragraph{\textbf{CF3 is not implied by CF1, CF2, CF4, and CF5.}}
Consider one round with
\[
	\bm d'_1=\bm a_1= (0.5,0.5), \qquad \text{and} \qquad \Delta \bm c_1 = (0.5,0.5).
\]
CF1 holds because each agent's permitted update interval is $[0,0.5]$.
For each agent, the other agent receives utility only $0.5 \le 1$, so the premise of CF2 is false.
CF4 holds because both agents receive their full demands, and the premise of CF5 is false because neither allocation exceeds $1$.
However, CF3 fails because $\Delta c_{1,1} + \Delta c_{2,1} = 1 > 0$.

\paragraph{\textbf{CF4 is not implied by CF1, CF2, CF3, and CF5.}}
Fix any $\varepsilon \in (0,0.5)$.
In round~1, let
\[
	\bm d'_1=\bm a_1= (1.5,0.5), \qquad \text{and} \qquad \Delta \bm c_1= (-0.5,0.5).
\]
The round-2 starting credits are therefore $\bm c_2= (-0.5,0.5)$.
All five conditions hold in round~1.
In particular, CF1 holds because the permitted update intervals are $[-0.5,0]$ and $[0,0.5]$.
For agent~2, agent~1 receives utility $1.5 > E_{-2} = 1$, so CF2 requires $\Delta c_{2,1} \ge 0.5$, which holds with equality.
The premise of CF2 is false for agent~1.
CF3 holds with equality.
Finally, CF4 and CF5 hold because both agents receive their full demands.

In round~2, let
\[
	\bm d'_2= (1,1.5), \qquad \bm a_2= (0.5-\varepsilon,1.5), \qquad \text{and} \qquad \Delta\bm c_2= (0.5,-0.5).
\]
CF1 holds: the permitted update intervals for agents~1 and~2 are $[0,0.5+\varepsilon]$ and $[-0.5,0]$,
respectively.
CF2 holds because agent~2 receives utility $1.5>E_{-1}=1$, requiring $\Delta c_{1,2} \ge 0.5$, which holds with equality; the premise of CF2 is false for agent~2.
CF3 also holds with equality.

The premise of CF5 is false for both agents, since
\[
	a_{1,2} = 0.5-\varepsilon < e_1+c_{1,2} = 0.5, \qquad \text{and} \qquad a_{2,2} = 1.5 = e_2 + c_{2,2}.
\]
However, CF4 fails for agent~1 because
\[
	\min( d'_{1,2}, e_1 + \min(0,c_{1,2}) ) = \min(1,0.5) = 0.5,
\]
whereas $a_{1,2} = 0.5 - \varepsilon < 0.5$.

\paragraph{\textbf{CF5 is not implied by CF1--CF4.}}
In round~1, let
\[
	\bm d'_1=\bm a_1= (0.5,1.5) \qquad \text{and}\qquad \Delta\bm c_1= (0.5,-0.5).
\]
The round-2 starting credits are therefore $\bm c_2= (0.5,-0.5)$.
All five conditions hold in round~1.
CF1 holds because the updates are at the endpoints of the permitted intervals.
CF2 holds because agent~2 receives utility $1.5>E_{-1}=1$ and agent~1 gains exactly $0.5$ credit.
CF3 holds with equality, and CF4 holds because both agents receive their full demands.
Although agent~2 receives strictly more than their credit-adjusted endowment of $1$, CF5 holds because both agents are fully satisfied.

In round~2, let
\[
	\bm d'_2= (1.5,1), \qquad \bm a_2=(1,1), \qquad \text{and}\qquad \Delta\bm c_2=(0,0).
\]
CF1 holds because both agents receive utility equal to their endowment.
The premise of CF2 is false for both agents because each other-agent utility equals, rather than exceeds, $1$.
CF3 holds.
CF4 also holds.
For agent~1, the required lower bound is
\[
	\min(1.5, 1+\min(0,0.5) ) = 1,
\]
and for agent~2 it is
\[
	\min(1, 1+\min(0,-0.5) ) = 0.5.
\]
Both bounds are respected.
However, CF5 fails.
Agent~2 receives
\[
	a_{2,2}=1 > \max(0,e_2+c_{2,2}) = 0.5.
\]
At the same time, agent~1 receives neither their full demand nor their credit-adjusted endowment:
\[
	a_{1,2}=1 < d'_{1,2} = 1.5, \qquad \text{and} \qquad a_{1,2} = 1 < e_1 + c_{1,2} = 1.5.
\]
\end{proof}

\subsection{Proof of Lemma~\ref{lem:pe-forces-delta}}\label{subsec:proof-of-lem:pe-forces-delta}

Since the total demand at round $t$ is at least as large as the total supply, Pareto efficiency implies that
\begin{equation}\label{eq:valued-sum-E}
	\sum_{j} u'_{j,t}(a_{j,t}) \;=\; E.
\end{equation}

Given (CF1), if $u'_{i,t}(a_{i,t}) \ge e_i$, then $\Delta c_{i,t}\le 0$, and if $u'_{i,t}(a_{i,t}) \le e_i$, then $\Delta c_{i,t}\ge 0$.
We claim first that for every agent $i$ with $u'_{i,t}(a_{i,t}) < e_i$, it must hold that $\Delta c_{i,t} = e_i - u'_{i,t}(a_{i,t})$.
Indeed, by Equation~\ref{eq:valued-sum-E}, we have:
\[
\sum_{j \neq i} u'_{j,t}(a_{j,t})= E - u'_{i,t}(a_{i,t}) > E - e_i	= E_{-i}.
\]
Therefore, the premise of (CF2) holds for agent $i$, which yields the following:
\begin{align*}
	\Delta c_{i,t} &\ge \sum_{j \neq i} u'_{j,t}(a_{j,t})- E_{-i}\\
	&= (E- u'_{i,t}(a_{i,t}))-(E-e_i)\\
	&= e_i - u'_{i,t}(a_{i,t}).
\end{align*}
Combined with the upper bound in (CF1), this implies that $\Delta c_{i,t}=e_i-u'_{i,t}(a_{i,t})$.

It remains to handle agents with $u'_{i,t}(a_{i,t}) \ge e_i$. Define set $S \triangleq \{i \in [n]: u'_{i,t}(a_{i,t}) \ge e_i\}$. For any agent~$i \in S$, the lower bound in (CF1) implies that $\Delta c_{i,t} \ge e_i - u'_{i,t}(a_{i,t})$. Summing $\Delta c_{j,t}$ over all agents, we obtain:
\begin{align*}
	\sum_{j \in [n]} \Delta c_{j,t} &= \sum_{j \in [n] \setminus S} \Delta c_{j,t} + \sum_{j \in S} \Delta c_{j,t} \\
	&\ge \sum_{j\in [n] \setminus S} (e_j - u'_{j,t}(a_{j,t})) + \sum_{j \in S} (e_j - u'_{j,t}(a_{j,t}))\\
	&= \sum_{j \in [n]} (e_j - u'_{j,t}(a_{j,t})) = 0,
\end{align*}
where the final equality follows from Equation~\ref{eq:valued-sum-E}. However, (CF3) requires $\sum_{j} \Delta c_{j,t} \le 0$, which enforces strict equality, implying that $\Delta c_{i,t} = e_i - u'_{i,t}(a_{i,t})$ for all agents $i$ with $u'_{i,t}(a_{i,t}) \ge e_i$.

We have shown that $\Delta c_{i,t} = e_i - u'_{i,t}(a_{i,t})$ for all agents $i$. When total demand is at least total supply, Pareto efficiency implies that $a_{i,t} \le d'_{i,t}$. Substituting $u'_{i,t}(a_{i,t})= a_{i,t}$ therefore yields the desired identity $\Delta c_{i,t} \;=\; e_i - a_{i,t}$. \hfill \qedsymbol

\subsection{Proof of Theorem~\ref{thm:CF-implies-SI}}\label{subsec:proof-of-thm:cf-implies-si}

The proof requires two lemmas. The first states that an agent's realized utility plus their current credit balance is always at least as large as the utility they would have obtained by simply keeping their own endowment in every round.

\begin{lemma}\label{lem:credit-adjusted-SI}
	Let $M$ be a credit-fair and Pareto efficient mechanism with an associated credit system $(c_{i,t})_{i \in [n], t \in \mathbb{N}}$. For every round $t$, every agent $i$, all true demands $\bm{d}_{i,1:t}$, and report profiles $\bm{d}'_{-i,1:t}$ of the other agents, it holds that:
	\[
	U_{i,t}\!\left(\bm{a}^M_{i,1:t}(\bm{d}_{i,1:t}, \bm{d}'_{-i,1:t})\right)+c_{i,t+1} \ge U_{i,t}(\bm{e}_i).
	\]
\end{lemma}
		
\begin{proof}
	For notational convenience, we suppress the distinction between true demands and reports and write $d_{j,t}$ for the demands observed by the mechanism.
	We will show that for every agent $i$ and every round $t \ge 1$, it holds that:
	\[
	u_{i,t}(a_{i,t}) + \Delta c_{i,t} \;\ge\; u_{i,t}(e_i).
	\]
	The lemma statement then follows directly by summing over $\tau = 1, \ldots, t$ and noting that $c_{i,1}=0$ by definition.

	We distinguish two cases depending on the total demand.

	\smallskip
	\noindent\textbf{Case A:} $\sum_{j} d_{j,t} < E$ (i.e., under-demand).
	By Pareto efficiency, no agent is under-served when there is slack, which implies that $u_{j,t}(a_{j,t}) = d_{j,t}$ for all $j$.
	Fix any agent $i$. If $d_{i,t} \le e_i$, then:
	\[
	u_{i,t}(e_i) = \min(d_{i,t}, e_i) = d_{i,t} = u_{i,t}(a_{i,t}).
	\]
	By (CF1), we have $\Delta c_{i,t} \ge 0$, and hence:
	\[
	u_{i,t}(a_{i,t}) + \Delta c_{i,t} \;\ge\; u_{i,t}(a_{i,t}) = u_{i,t}(e_i).
	\]
	If $d_{i,t} > e_i$, then $u_{i,t}(e_i) = e_i$, and $u_{i,t}(a_{i,t}) = d_{i,t} > e_i$.
	By (CF1), we have: $e_i - u_{i,t}(a_{i,t}) \le \Delta c_{i,t} \le 0$, and therefore:
	\[
	u_{i,t}(a_{i,t}) + \Delta c_{i,t} \;\ge\; e_i = u_{i,t}(e_i).
	\]
	Thus, the lemma holds in Case A\@.

	\smallskip
	\noindent\textbf{Case B:} $\sum_{j} d_{j,t} \ge E$ (i.e., shortage).
	By Pareto efficiency, the full supply is allocated, and no agent receives more resources than their demand.
	Therefore, $\sum_{j} a_{j,t} = E$, and for all $j$, $u_{j,t}(a_{j,t}) = a_{j,t}$ (equivalently, $d_{j,t} \ge a_{j,t}$).
	Fix any agent $i$. Substituting $\Delta c_{i,t}=e_i-a_{i,t}$ from Lemma~\ref{lem:pe-forces-delta} gives the following:
	\[
	u_{i,t}(a_{i,t})+\Delta c_{i,t} = a_{i,t}+\Delta c_{i,t} = e_i \ge \min(d_{i,t}, e_i) = u_{i,t}(e_i).
	\]
	Thus, the lemma holds in Case B as well.
\end{proof}
		
We next identify the rounds in which an agent can fall below their per-round baseline, and how this interacts with their credit balance.

\begin{lemma}\label{lem:bad-rounds}
	Let $M$ be a credit-fair and Pareto efficient mechanism with associated credit system $(c_{i,t})_{i \in [n], t \in \mathbb{N}}$. Fix an agent $i$ and a round $t$ such that $u'_{i,t}(a_{i,t}) < u'_{i,t}(e_i)$.
	Then, it must be that (1) $c_{i,t} < 0$, and (2) $c_{i,t+1} \le 0$.
\end{lemma}
		
\begin{proof}
	As in the proof of Lemma~\ref{lem:credit-adjusted-SI}, we again suppress the distinction between true demands and reports (and therefore between $u'_{i,t}$ and $u_{i,t}$).
	Suppose $u_{i,t}(a_{i,t}) < u_{i,t}(e_i)$.
	First, if $\sum_{j} d_{j,t} < E$, then, as in the proof of Lemma~\ref{lem:credit-adjusted-SI}, Pareto efficiency implies that $a_{j,t} \ge
	d_{j,t}$ for all $j$.
	Hence, $u_{i,t}(a_{i,t}) = d_{i,t} \ge \min (d_{i,t}, e_i)=u_{i,t}(e_i)$, which contradicts the assumption.
	Thus, it must be that $\sum_{j} d_{j,t} \ge E$; that is, round $t$ is a shortage round.

	We first show that $c_{i,t} < 0$.
	In a shortage round, $u_{i,t}(a_{i,t}) = a_{i,t}$ and $u_{i,t}(e_i) =
	\min(d_{i,t},e_i)$. The inequality $u_{i,t}(a_{i,t}) < u_{i,t}(e_i)$ translates
	to $a_{i,t} < \min(d_{i,t},e_i)$. If $c_{i,t} \ge 0$, then (CF4) implies that:
	\[
	a_{i,t} \;\ge\; \min(d_{i,t},\, e_i + \min(0,c_{i,t})) = \min(d_{i,t}, e_i),
	\]
	a contradiction.

	We next show that $c_{i,t+1} \le 0$.
	Consider a shortage round with $a_{i,t} < \min(d_{i,t},e_i)$ and $c_{i,t} <
	0$. This implies that $a_{i,t} < e_i$ and $u_{i,t}(a_{i,t}) = a_{i,t}$.
	By (CF4), it must hold that:
	\[
	a_{i,t} \;\ge\; \min(d_{i,t},\, e_i + \min(0,c_{i,t})) = \min(d_{i,t},\, e_i + c_{i,t}).
	\]
	As in Case B of Lemma~\ref{lem:credit-adjusted-SI}, shortage together with
	$a_{i,t} < e_i$ implies, via (CF1) and (CF2), that $\Delta c_{i,t} = e_i - a_{i,t}$.
	Moreover, from $a_{i,t} \ge \min(d_{i,t}, e_i + c_{i,t})$ and the fact that $a_{i,t} < \min(d_{i,t},e_i)$, it follows that $a_{i,t} \ge e_i + c_{i,t}$. Hence, we have:
	\[
	\Delta c_{i,t} = e_i - a_{i,t} \;\le\; -c_{i,t}.
	\]
	Therefore, $c_{i,t+1} = c_{i,t} + \Delta c_{i,t} \le 0$.
\end{proof}

We can now show that sharing incentives holds.

\begin{proof}[Proof of Theorem~\ref{thm:CF-implies-SI}]
	Fix an agent $i$, a demand sequence $\bm{d}_{i,1:t}$, a report profile $\bm{d}'_{-i,1:t}$ of the other agents, and a round $t \ge 1$.
	We need to show that $U_{i,t}(\bm{a}_{i,1:t}) \ge U_{i,t}(\bm{e}_i)$ for all $t$.
	Suppose, for contradiction, that there exists a round $t$ such that $U_{i,t}(\bm{a}_{i,1:t}) < U_{i,t}(\bm{e}_i)$.
	Let $t^\star$ be the smallest such round.
	Then $U_{i,t^\star}(\bm{a}_{i,1:t^\star}) - U_{i,t^\star}(\bm{e}_i) < 0$ and $U_{i,t^\star-1}(\bm{a}_{i,1:t^\star-1}) - U_{i,t^\star-1}(\bm{e}_i) \ge 0$.
	In particular, $u_{i,t^\star}(a_{i,t^\star}) - u_{i,t^\star}(e_i) < 0$.
	Therefore, the condition of Lemma~\ref{lem:bad-rounds} is satisfied for round $t^\star$ and agent $i$.
	By that lemma we have $c_{i,t^\star+1} \le 0$.

	On the other hand, Lemma~\ref{lem:credit-adjusted-SI} applied at round $t^\star$ gives
	\[
	U_{i,t^\star}(\bm{a}_{i,1:t^\star}) + c_{i,t^\star+1} \;\ge\; U_{i,t^\star}(\bm{e}_i).
	\]
	Equivalently,
	\[
	U_{i,t^\star}(\bm{a}_{i,1:t^\star}) - U_{i,t^\star}(\bm{e}_i) \;\ge\; -\,c_{i,t^\star+1} \;\ge\; 0,
	\]
	where the last inequality follows from $c_{i,t^\star+1} \le 0$.
	This contradicts the assumption that $U_{i,t}(\bm{a}_{i,1:t}) < U_{i,t}(\bm{e}_i)$ for some $t \ge 1$.
\end{proof}

\subsection{Proof of Proposition~\ref{prop:smm-not-credit-fair}}

Consider $n=2$ agents, each with endowment $e_i = 1$, and the first $2$ rounds with the following demands:
{\footnotesize
	\[
		 \begin{array}{c|cccc}
			 & \text{$t=1$} & \text{$t=2$} \\
			 \hline
			 \text{Agent 1} & 0 & 2 \\
			 \text{Agent 2} & 2 & 2 \\
		 \end{array}
	\]
}
SMMF yields the following allocations:
{\footnotesize
	\[
		\begin{array}{c|cccc}
			& \text{$t=1$} & \text{$t=2$} \\
			\hline
			\text{Agent 1} & 0 & 1 \\
			\text{Agent 2} & 2 & 1 \\
		\end{array}
	\]
}
If SMMF were credit fair with credit system $(c_{i,t})_{i \in [n], t \in \mathbb{N}}$, then after round~1 the following must hold:
\begin{itemize}
	\item Agent 1 donates one unit in round 1, therefore, by (CF1) and (CF2), we have $c_{1,2} = 1$.
	\item By (CF3), it follows that $c_{2,2} \le -1$.
\end{itemize}

Since $a_{2,2} = 1 > \max(0, e_2 + c_{2,2}) = 0$, credit fairness
Condition (CF5) applies with $j = 2$ and requires that for agent~1 either
$a_{1,2} = d_{1,2}$ or $a_{1,2} \ge e_1 + c_{1,2} = 2$, both of which are false. \hfill\qedsymbol

\subsection{Static Mechanism Is Credit Fair}\label{subsec:static-mechanism-is-credit-fair}

We provide a proof that the static mechanism that always allocates every agent their endowment $a_{i,t}=e_i$ is credit fair.

\begin{proposition}\label{prop:endowment-credit-fair}
	Let $M^{\text{static}}$ be the mechanism that, in every round $t$ and for every
	report profile, allocates each agent exactly their endowment:
	\[
	a_{i,t} \;=\; e_i \qquad \text{for all } i \in [n],\ t\ge 1.
	\]
	Then $M^{\text{static}}$ is credit fair.
\end{proposition}

\begin{proof}
	We exhibit a credit system $(c_{i,t})_{i \in [n], t \in \mathbb{N}}$ satisfying conditions
	(CF1)--(CF5). Define
	\[
	c_{i,t} \coloneqq 0 \qquad \text{for all } i\in[n],\ t\ge 1,
	\]
	so that $\Delta c_{i,t} = c_{i,t+1}-c_{i,t}=0$ for all $i,t$.

	We verify each condition.

	\smallskip
	\noindent\textbf{(CF1).}
	Since $a_{i,t}=e_i$, we have $u'_{i,t}(a_{i,t})=u'_{i,t}(e_i)$, which implies:
	\[
	e_i-u'_{i,t}(a_{i,t})= e_i-u'_{i,t}(e_i) \ge 0.
	\]
	(CF1) requires the following:
	\[
	\min\!\bigl(0,\,e_i- u'_{i,t}(a_{i,t})\bigr) \;\le\; \Delta c_{i,t} \;\le\; \max\!\bigl(0,\,e_i- u'_{i,t}(a_{i,t})\bigr).
	\]
	The left-hand side equals $0$, and the right-hand side is nonnegative.
	Therefore, $\Delta c_{i,t}=0$ satisfies (CF1).

	\smallskip
	\noindent\textbf{(CF2).}
	Fix an agent $i$ and a round $t$. The premise of (CF2) is
	\[
	\sum_{j\neq i} u'_{j,t}(a_{j,t}) \;>\; E_{-i}.
	\]
	However, we have $a_{j,t}=e_j$, which implies $u'_{j,t}(a_{j,t})=u'_{j,t}(e_j)\le e_j$.
	Therefore, we have:
	\[
	\sum_{j\neq i}u'_{j,t}(a_{j,t}) \;\le\; \sum_{j\neq i} e_j \;=\; E_{-i}.
	\]
	Hence, the premise can never hold, so (CF2) is satisfied vacuously.

	\smallskip
	\noindent\textbf{(CF3).}
	We have $\sum_{i} \Delta c_{i,t} = \sum_{i} 0 = 0 \le 0$, so (CF3) holds.

	\smallskip
	\noindent\textbf{(CF4).}
	Since $c_{i,t}=0$, we have $\min(0,c_{i,t})=0$, and thus:
	\[
	\min\!\bigl(d'_{i,t},\,e_i+\min(0,c_{i,t})\bigr) = \min(d'_{i,t},e_i)\le e_i.
	\]
	Because $a_{i,t}=e_i$, we obtain:
	\[
	a_{i,t} = e_i \;\ge\; \min(d'_{i,t},e_i) = \min\!\bigl(d'_{i,t},\,e_i+\min(0,c_{i,t})\bigr),
	\]
	so (CF4) holds.

	\smallskip
	\noindent\textbf{(CF5).}
	Since $c_{j,t}=0$, the threshold in (CF5) is $\max(0,e_j+c_{j,t})=\max(0,e_j)=e_j$.
	But $a_{j,t}=e_j$, so the antecedent $a_{j,t}>\max(0,e_j+c_{j,t})$ is never satisfied. Thus (CF5) holds vacuously.

	\smallskip
	All five conditions are satisfied, so $M^{\text{static}}$ is credit fair.
\end{proof}

\section{Omitted Material from Section~\ref{sec:algorithm}} \label{sec:app-algorithm}

\subsection{Proof of Theorem~\ref{thm:mech-osp}}

The following lemma regarding the behavior of the \texttt{PSWC} procedure will be helpful; the proof is immediate from the definition of PSWC.

\begin{lemma} \label{lem:pswc-monotonicity} 
	Fix a PSWC instance with resource capacity $A$, weights $\bm{w}$, lower bounds $\bm{m}$, and upper bounds $\bm{l}$, and let $\bm{a}$ be its output. Fix an agent $i$, and consider a second feasible PSWC instance obtained by changing only agent $i$'s bounds, with output $\bm{a}'$. Then: \begin{enumerate} \item increasing either $m_i$ or $l_i$, while holding all other parameters fixed, cannot decrease agent $i$'s allocation; and \item decreasing either $m_i$ or $l_i$, while holding all other parameters fixed, cannot increase agent $i$'s allocation. \end{enumerate} Furthermore, if $m_i < a_i < l_i$, then changing agent $i$'s bounds while keeping $a_i$ within the new bounds leaves agent $i$'s allocation unchanged. 
\end{lemma} 

We can now prove Theorem~\ref{thm:mech-osp}.

\begin{proof}[Proof of Theorem~\ref{thm:mech-osp}]
	Fix a round $t \ge 1$, an agent $i$ with reports $\bm{d}'_{i,1:t-1}=\bm{d}_{i,1:t-1}$ in rounds $1, \ldots, t-1$. Let $\bm{d}'_{-i,1:t}$ denote the report profile of all the other agents and write $\bm{d}'_{-i,t}$ for their reports in round $t$. Finally, let $d \triangleq d_{i,t}$ denote agent $i$'s true demand in round $t$.

	The history up to round $t-1$ fixes the credit vector $\bm c_t$ and the past allocations $(\bm a_1,\dots,\bm a_{t-1})$, hence fixes the quantities:
	\[
	A^{\mathrm{past}}_j \;:=\; \sum_{t'=1}^{t-1} a_{j,t'}
	\]
	for all $j \in [n]$.
	Therefore, the only effect of any misreport by agent $i$ in round $t$ is through their report $r \triangleq d'_{i,t}$, which enters the round-$t$ branch conditions and the \texttt{PSWC} constraints.

	Write $S \triangleq \sum_{j\ne i} d'_{j,t}$, so the total reported demand at round $t$ is $D(r) \triangleq r+S$.
	We compare agent $i$'s utility in round $t$ under truthful reporting $r=d$ with that under an
	arbitrary misreport $r\ge 0$.
	Throughout this proof, we treat several quantities as functions of the report in order to clarify the effect of agent $i$'s report; for example, we write $a_{i,t}(d)$ and $a_{i,t}(r)$ for agent $i$'s allocation in round $t$ under reports $d$ and $r$, respectively, holding all other reports fixed.

	\medskip
	\noindent\textbf{Case 1: $D(d) \le E$ (truth is in the no-shortage branch).}
	When $D(d)\le E$, $\mech$ calls \texttt{PSWC} with minima $\bm m=\bm{d}'_t$ and no upper limits.
	In particular, under truthful reporting we have:
	\[
	a_{i,t}(d)\;\ge\; d,
	\]
	so $u_{i,t}(a_{i,t}(d))=\min(d,a_{i,t}(d))=d$, which is the maximum possible utility in round $t$.
	Therefore, no deviation $r$ can strictly improve utility.

	\medskip
	\noindent\textbf{Case 2: $D(d)>E$ (truth is in a shortage branch).}
	Define the \emph{credit-adjusted cap} (fixed under the history):
	\[
	\kappa_i \triangleq \max(0,e_i+c_{i,t}),
	\]
	and the corresponding capped report $\bar d_i(r) \triangleq \min(r,\kappa_i)$.

	We split by whether the deviation changes the outer branch, i.e., from shortage to no shortage.

	\smallskip
	\noindent\emph{Case 2a: Deviation chooses $r$ with $D(r)\le E$ (switching to no-shortage).}
	Define $r'$ as follows:
	\[
	r' \;:=\; r + (E-D(r)) \;=\; E-\sum_{j\ne i} d_{j,t},
	\]
	so that $D(r')=E$.
	We first show that the deviation $r'$ weakly dominates $r$ in round $t$, and hence it suffices to analyze the boundary deviation with $D(r')=E$.

	In the no-shortage branch, $\mech$ runs \texttt{PSWC} with total resource $A=E$, weights $\bm w=\bm e$, minimum constraints $\bm m=\bm d'_t$, and no upper limits.
	Thus, every agent $j$ receives at least $m_j=d'_{j,t}$, and the total resource left after satisfying all minima is $E-D(r)$.
	Consequently, agent $i$'s allocation in round $t$ satisfies the following:
	\begin{equation}\label{eq:case2a-upper}
		a_{i,t}(r)\;\le\; r + (E-D(r)) \;=\; r'.
	\end{equation}
	Now consider the deviation $r'$. Since $\sum_j m_j = D(r')=E$, feasibility
	forces \texttt{PSWC} to return $a_{j,t}(r')=m_j=d'_{j,t}$ for every agent $j$, and in particular $a_{i,t}(r')=r'$.
	Combining with \eqref{eq:case2a-upper} and using that $u_{i,t}$ is nondecreasing in the allocation, we obtain:
	\[
	u_{i,t}\!\left(a_{i,t}(r)\right) \;\le\; u_{i,t}\!\left(a_{i,t}(r')\right).
	\]
	Therefore, among deviations that switch to the no-shortage branch, it suffices to consider $r'$ with $D(r')=E$.

	It remains to compare truthful reporting with the deviation $r'$.
	Since $D(d)>E$ by assumption, truthful reporting lies in the shortage branch.
	As shown in Case 2b below, no deviation that remains in the shortage branch, and in particular, no deviation with $D(r')=E$, can yield strictly higher utility in round $t$ than truthful reporting.
	(Note that we can treat reports with $D(r')=E$ as lying in either the shortage or no-shortage branches, since this case has a unique Pareto efficient allocation that will be reached by either branch of the algorithm).
	Hence, no deviation that switches to the no-shortage branch can be profitable.

	\smallskip
	\noindent\emph{Case 2b: Deviation chooses r with $D(r)>E$ (staying in shortage).}
	Now both truthful reporting and the deviation remain in the shortage branch.
	There are two subbranches depending on whether $E\le \sum_j \bar d_j(d)$.
	In all cases, it must hold that $a_{i,t}(r) \le r$ since the deviation is in a shortage branch.

	\smallskip
	\noindent\emph{Subcase 2b(i): $E\le \sum_j \bar d_j(d)$ under truthful reporting.}
	We first argue that any deviation $r$ that changes the \emph{inner} shortage
	branch, i.e., that makes $\sum_j \bar d_j(r) < E$, cannot be beneficial.

	Since only agent $i$ deviates, for all $j\neq i$ we have $\bar d_j(r)=\bar d_j(d)$.
	Therefore, $\sum_j \bar d_j(r) < E \le \sum_j \bar d_j(d)$ can hold only if $\bar d_i(r) < \bar d_i(d)$.
	In particular, this implies $r<\kappa_i$, and hence $\bar d_i(r)=r$.
	Moreover, we have:
	\[
	\sum_j \bar d_j(r) < E
	\quad\Longleftrightarrow\quad
	r < E-\sum_{j\neq i}\bar d_j(d).
	\]
	Next consider truthful reporting in this subcase.
	Under truthful reporting, $\mech$ runs the \texttt{PSWC} call with upper limits $\bar{\bm d}(d)$ and allocates total
	resource $E$.
	Since every other agent $j\neq i$ receives at most $\bar d_j(d)$ in this call, the feasibility of allocating $E$ implies the following:
	\[
	a_{i,t}(d)\;\ge\; E-\sum_{j\neq i}\bar d_j(d)\;>\;r \ge a_{i,t}(r),
	\]
	and the misreport is not helpful (by monotonicity of $u_{i,t}$).

	Now suppose that $\sum_j \bar{d}_j(r) \ge E$. Then, under truthful reporting, $\mech$ runs \texttt{PSWC} with minima $\bm m=\bm 0$ and upper limits $\bm l = \bar{\bm d}(d)$.
	Agent $i$'s report affects this \texttt{PSWC} call only through the upper limit $l_i=\bar{d}_i(r)$.

If under truthful reporting $a_{i,t}(d) < l_i(d)$, then the upper bound is nonbinding, so by Lemma~\ref{lem:pswc-monotonicity}, increasing $l_i$ by overreporting leaves $a_{i,t}$ unchanged, while decreasing $l_i$ by underreporting cannot increase $a_{i,t}$. Hence $a_{i,t}(r) \le a_{i,t}(d)$ and therefore utility cannot increase.

	If instead, the upper bound binds under truthful reporting, i.e.\ $a_{i,t}(d)=l_i(d)=\bar
	d_i(d)$, then either $\bar{d}_i(d)=d$ (in which case $u_{i,t}(a_{i,t}(d))=d$ is
	maximal) or $\bar{d}_i(d)=\kappa_i \le d$ (in which case $\bar{d}_i(r)\le \kappa_i$ for
	all $r$ and thus no deviation can increase $a_{i,t}$ beyond $\kappa_i$).
	Either way, no deviation increases agent $i$'s utility in round $t$.

	\smallskip
	\noindent\emph{Subcase 2b(ii): $E> \sum_j \bar d_j(d)$ under truthful reporting.}
	We first argue that any deviation $r$ that flips the inner shortage branch in the
	opposite direction, i.e., that makes $E\le \sum_j \bar d_j(r)$, cannot be
	beneficial for agent $i$.

	Such a flip necessarily requires $\bar d_i(r)>\bar d_i(d)$, hence $r>d$ and $d<\kappa_i$.
	Using only the latter inequality, we know that $\bar{d}_i(d)=d$ and so under truthful reporting, \mech runs \texttt{PSWC} with
	\[
	m_i(d)=A^{\mathrm{past}}_i+d, \qquad l_i(d)=A^{\mathrm{past}}_i+d,
	\]
	and then sets $a_{i,t}(d)=\tilde a_{i,t}(d)-A^{\mathrm{past}}_i=d$, so that $u_{i,t}(a_{i,t}(d)) = d$ is maximal.

	Finally, consider a misreport that does not change the inner branch. If $d\le \kappa_i$ then the same argument as above holds, so suppose that $d > \kappa_i$ (i.e., $\bar{d}_i(d)<d$). 
	Any underreport $r < d$ can only reduce the bounds $m_i$ and $l_i$. Therefore, by Lemma~\ref{lem:pswc-monotonicity}, it can only result in $a_{i,t}(r) \le a_{i,t}(d)$, and hence $u_{i,t}(a_{i,t}(r)) \le u_{i,t}(a_{i,t}(d))$.
	An overreport $r > d$ affects only the upper limit $l_i$, with $l_i(r) > l_i(d)$. By Lemma~\ref{lem:pswc-monotonicity}, this leaves agent $i$'s allocation unchanged if the upper bound is nonbinding under truthful reporting, and can only increase the allocation otherwise. Thus, either $a_{i,t}(r) = a_{i,t}(d)$ if $\tilde{a}_{i,t}(d) < l_i(d)$, or $a_{i,t}(r) \ge a_{i,t}(d)$ if the upper bound binds under truthful reporting. In the latter case, $a_{i,t}(d)=d$, so $u_{i,t}(a_{i,t}(r)) = d = u_{i,t}(a_{i,t}(d))$.

	In all subcases, deviations that remain in the shortage branch cannot increase $i$'s utility in round $t$.

	\medskip
	Combining Cases 1 and 2, we conclude that for every deviation $r\ge 0$,
	\[
	u_{i,t}(a_{i,t}(d)) \;\ge\; u_{i,t}(a_{i,t}(r)),
	\]
	which establishes online strategy-proofness.
\end{proof}

\end{document}